\documentclass[symmetry,article,submit,moreauthors,pdftex]{mdpi} 
\pdfoutput=1

\renewcommand{\linenumbers}{}

\firstpage{1} 
\pubvolume{1}
\issuenum{1}
\articlenumber{0}
\pubyear{2022}
\copyrightyear{2021}
\datereceived{} 
\dateaccepted{} 
\datepublished{} 
\hreflink{https://doi.org/} % If needed use \linebreak

\usepackage{graphicx}
\usepackage{tabularx}
\usepackage{bm}
\usepackage{color}
\definecolor{darkblue}{RGB}{0,0,196}
\definecolor{darkred}{RGB}{196,0,0}

\def\be{\begin{equation}}
\def\ee{\end{equation}}
\def\ba{\begin{eqnarray}}
\def\ea{\end{eqnarray}}

\Title{Resummed relativistic dissipative hydrodynamics}

\TitleCitation{Resummed relativistic dissipative hydrodynamics}

\Author{Huda Alalawi $^{1}$\orcidA{},  Mubarak Alqahtani $^{2}$\orcidB{}, and Michael Strickland $^{1}$\orcidC{}}

\AuthorNames{Huda Alalawi, Mubarak Alqahtani, Michael Strickland}

\AuthorCitation{Alalawi, Huda, Mubarak Alqahtani, and Michael Strickland}
\address{%
$^{1}$ \quad Department of Physics, Kent State University, Kent, OH 44242 United States; halalawi@kent.edu and mstrick6@kent.edu \\
$^{2}$ \quad Department of Basic Sciences, College of Education, Imam Abdulrahman Bin Faisal University, Dammam 34212, Saudi Arabia; maqahtani@iau.edu.sa}

\abstract{In this review, we present the motivation for using relativistic anisotropic hydrodynamics to study the physics of ultrarelativistic heavy-ion collisions.  We then highlight the main ingredients of the 3+1D quasiparticle anisotropic hydrodynamics model and present phenomenological comparisons with experimental data at different collision energies. These comparisons show that anisotropic hydrodynamics can describe many bulk observables of the quark-gluon plasma.}


\begin{document}
%%%%%%%%%%%%%%%%%%%%%%%%%%%%%%%%%%%%%%%%%%

%%%%%%%%%%%%%%%%%%%%%%%%%%%%%%%%%%%%%%%%%%
\section{Introduction}
\label{sec:intro}
%%%%%%%%%%%%%%%%%%%%%%%%%%%%%%%%%%%%%%%%%%

Experiments at the Relativistic Heavy Ion Collider (RHIC) at Brookhaven National Laboratory and the Large Hadron Collider (LHC) at CERN are probing the nature of hot and dense matter by colliding heavy-ions at ultrarelativistic center of mass energies of up to 5 TeV per nucleon \cite{Averbeck2015,Wang:2016opj}.
The goal of these experiments is to generate conditions similar to those present in the early universe and during mergers of compact astrophysical objects such as neutron stars \cite{Dexheimer:2020zzs}.
These conditions correspond to high temperature ($T \gtrsim 150$ MeV) and net baryon density ($\rho \gtrsim 2-3\,\rho_{\rm sat}$), respectively.
In both cases it is expected that nuclear matter undergoes a phase transition from a state in which quarks and gluons are confined inside hadrons to a deconfined state, called the quark-gluon plasma (QGP), in which quarks and gluons are not bound inside of hadrons.
At finite temperatures and zero net baryon density it is possible to make use of lattice quantum chromodynamics (QCD) to determine the temperature at which the deconfinement transition occurs and the nature of the transition.
For realistic quark masses, continuum extrapolated lattice QCD calculations find that the transition is a smooth crossover with a pseudocritical temperature $T_{\rm pc} \simeq 155$ MeV \cite{Bazavov:2013txa,Borsanyi:2016bzg}.
Due to the fermionic sign problem it is not possible to perform calculations at finite baryochemical potential $\mu_B$, however, it is possible to make use of Taylor expansions around $\mu_B = 0$ or analytic continuations of imaginary chemical potential calculations to determine quantities of interest such as various quark susceptibilities and the curvature of the QCD phase transition line itself \cite{Cea:2015cya,Bonati:2015bha,Bonati:2018nut,Bonati:2014rfa,Borsanyi:2020fev,Bazavov:2018mes,Toublan:2004ks,Endrodi:2011gv,Bellwied:2015rza,2006,Haque:2020eyj}.  These measurements provide constraints on the equation of state of QCD which can then be used in dynamical simulations of QGP evolution.

For modeling the spatiotemporal dynamics of the QGP created in ultrarelativistic heavy-ion collisions, one of the main tools used is relativistic viscous hydrodynamics \cite{Jeon:2016uym,2010,Romatschke:2017ejr}.  Early studies using relativistic hydrodynamics used the ideal limit \cite{Huovinen_2001,Hirano:2002ds,Kolb:2003dz} in which all dissipative transport coefficients, such as the shear viscosity were assumed to be zero, however, this was known to be an idealization because one expects, both on quantum uncertainty and causality bounds, that the ratio of the shear viscosity to entropy density ratio should have a lower bound.  In order to incorporate such dissipative transport coefficients in the dynamics it was necessary to develop a causal version of viscous hydrodynamics called {\em second-order viscous hydrodynamics} \cite{Muronga:2001zk,Muronga:2003ta,Muronga:2004sf,Heinz:2005bw,Baier:2006um,Romatschke:2007mq,Baier:2007ix,Denicol:2012cn,Denicol:2012es,Jaiswal:2013npa,Jaiswal:2013vta,Calzetta:2014hra,Denicol:2014vaa,Denicol:2014mca,Jaiswal:2014isa}.  Application of second-order viscous hydrodynamics to QGP phenomenology quickly followed, with practitioners able to extract estimates of the shear viscosity to entropy density ratio which were consistent with the generation of a strongly-coupled QGP.  Progress since then has included the development of consistent second-order truncations of the relativistic dissipative hydrodynamics from relativistic kinetic theory \cite{Denicol:2012cn,Denicol:2012es} and recently formulations of casual first-order formulations which make use of different hydrodynamic frames \cite{Bemfica:2020zjp,Hoult:2020eho}.

One of the major issues faced by second-order formulations of dissipative relativistic hydrodynamics is that, at very early times after the nuclear pass through, the system is quite far from equilibrium, with the largest non-equilibrium deviations reflected in the fact that, in the local rest frame (LRF), the system possesses a much smaller pressure along the beam-line direction (longitudinal direction) than transverse to it, i.e. $P_L \ll P_T$.  This LRF pressure anisotropy emerges due to the rapid longitudinal expansion of the QGP and has been shown to exist in both the weak- and strong-coupling limits, with the anisotropy becoming more pronounced as the coupling is decreased.  The implication of this is that the viscous corrections, in particular the shear correction, are large at early times, calling into doubt the reliability of fixed-order truncations in the magnitude of the inverse Reynolds number \cite{Heinz:2014zha}.

Another issue faced by fixed order truncations of viscous hydrodynamics is that, due to the assumed polynomial from of the corrections to the one-particle distribution function, there is the possibility that the viscous-corrected one-particle distribution function can become negative, which violates the positivity of probabilities.  In order to address both of these issues, in {\em anisotropic hydrodynamics} (aHydro) one makes use of a form for the one-particle distribution function that is, by construction, non-negative while also having kinetic pressures which are non-negative.  The original papers on aHydro focused on application to systems undergoing boost-invariant conformal Bjorken expansion \cite{Florkowski:2010cf,Martinez:2010sc}.  In Ref.~\cite{Martinez:2010sc} it was demonstrated that one could obtain both the ideal hydrodynamics and free-streaming limits in the aHydro framework and numerical solutions to the resulting coupled evolution equations demonstrated that both the one-particle distribution function and the kinetic pressures remained positive at all times.  Since then many works have extended these initial studies to include more realistic features associated with heavy-ion collisions, ultimately allowing practitioners to simulate the full three-dimensional non-conformal evolution of the QGP with a lattice-based equation of state \cite{Ryblewski:2010ch,Florkowski:2011jg,Martinez:2012tu,Ryblewski:2012rr,Bazow:2013ifa,Tinti:2013vba,Nopoush:2014pfa,Tinti:2015xwa,Bazow:2015cha,Strickland:2015utc,Alqahtani:2015qja,Molnar:2016vvu,Molnar:2016gwq,Alqahtani:2016rth,Bluhm:2015raa,Bluhm:2015bzi,Alqahtani:2017jwl,Alqahtani:2017tnq,Almaalol:2018jmz,Almaalol:2018ynx,McNelis:2018jho,Nopoush:2019vqc,McNelis:2021zji}.

In this review we will summarize the progress made in recent years including phenomenological applications.  We will begin with a demonstration that the aHydro dynamical equations resum an infinite series of terms when expanded as a power series in the inverse Reynolds number.  We will then present a review of the underpinnings of the 3+1D quasiparticle aHydro (aHydroQP) framework, which goes beyond traditional approaches by resumming viscous contributions to all orders in the shear and bulk inverse Reynolds numbers.  In this second part, we will focus on recent phenomenological applications of 3+1D aHydroQP to AA collisions at RHIC and LHC energies.

The structure of this review is as follows.  In Sec.~\ref{sec:resum} we discuss the case of conformal Bjorken expansion in order to demonstrate how anisotropic hydrodynamics resums contributions to all orders in the inverse shear Reynolds number.  In Sec.~\ref{sec:aHydroQP}, we introduce quasiparticle anisotropic hydrodynamics. In Sec.~\ref{sec:EoS}, we outline the construction of the QCD equation in aHydroQP. In Sec.~\ref{sec:freezeout}, evolution and freezeout are discussed in the 3+1D aHydroQP model. In Sec.~\ref{sec:Results}, phenomenological comparisons to experimental data are presented at various collision energies. Sec.~\ref{sec:conclusions} contains our conclusions and a summary of ongoing projects. 

%%%%%%%%%%%%%%%%%%%%%%%%%%%%%%%%%%%%%%%%%%
\section{Resummed dissipative hydrodynamics in the conformal Bjorken limit}
\label{sec:resum}
%%%%%%%%%%%%%%%%%%%%%%%%%%%%%%%%%%%%%%%%%%

Before presenting the full 3+1D formalism for non-conformal QCD plasmas, it is instructive to consider the 0+1D conformal limit in which the system undergoes Bjorken expansion.
In the case of conformal Bjorken expansion, at zero chemical potential, the aHydro distribution function contains a single independent anisotropy parameter $\xi$.  In the local rest frame of the plasma the distribution takes the form \cite{Romatschke:2003ms}
\be
f(x,p) = f_{\rm eq}\left( \frac{\sqrt{{\bf p}^2 + \xi p_z^2}}{\lambda}\right) ,
\label{eq:rsform}
\ee
where $\lambda$ is a non-equilibrium momentum scale that becomes the temperature in the limit $\xi \rightarrow 0$ and $f_{\rm eq}$ is either a Boltzmann, Bose, or Fermi-Dirac distribution depending on the assumed equilibrium statistics of the particle being considered.

The evolution equation obtained from requiring energy-momentum conservation can be written compactly as
\be
\label{eq:energydens}
\frac{\partial {\epsilon (\tau)}}{\partial \tau}=-\frac{\epsilon (\tau)+P_L(\tau)}{\tau} \, ,
\ee
where $\epsilon$ is the energy density, $P_L$ is the longitudional pressure, and $\tau$ is the proper time in Milne coordinates.
Using \eqref{eq:rsform} this becomes \cite{Martinez:2010sc}
\be
\frac{{\cal R}'(\xi)}{{\cal R}(\xi)} \partial_\tau \xi + \frac{4}{\lambda} \partial_\tau \lambda = 
\frac{1}{\tau} \left[ \frac{1}{\xi(1+\xi){\cal R}(\xi)} - \frac{1}{\xi} - 1 \right] ,
\label{eq:firstmoment}
\ee
with 
\be
{\cal R}(\xi) = \frac{1}{2}\left[\frac{1}{1+\xi}
+\frac{\arctan\sqrt{\xi}}{\sqrt{\xi}} \right] .
\ee

In the relaxation time approximation, the second evolution equation required can be obtained from the second moment of the Boltzmann equation.  Following the Florkowski-Tinti prescription \cite{Tinti:2013vba} one obtains
\be
\frac{1}{1+\xi} \partial_\tau\xi - \frac{2}{\tau} + \frac{{\cal R}^{5/4}(\xi)}{\tau_{\rm eq}} \xi \sqrt{1+\xi} = 0\, .
\label{eq:2ndmomf}
\ee
where, for a conformal system, one has $\tau_{\rm eq} = 5 \bar\eta/T$ with $\bar\eta = \eta/s$ being the specific shear viscosity and the effective temperature $T = {\cal R}^{1/4}(\xi) \lambda$ determined by Landau matching.  In the conformal limit, the first and second moment equations are independent of the assumed form of $f_{\rm eq}$.  The final evolution equations \eqref{eq:firstmoment} and \eqref{eq:2ndmomf} for $\xi$ and $\lambda$ are highly non-linear but can be easily solved numerically.  We note that these equations reproduce the ideal hydrodynamic limit when $\tau_{\rm eq} \rightarrow 0$ and the free streaming limit when $\tau_{\rm eq} \rightarrow \infty$.

\subsection{Relation to second-order viscous hydrodynamics in the small anisotropy limit}
\label{sec:smallaniso}

In order to make a connection to standard second-order viscous hydrodynamics, one can rewrite Eqs.~(\ref{eq:energydens})~and~(\ref{eq:2ndmomf}) in terms of the single shear stress tensor component $\pi \equiv {\pi^\varsigma}_\varsigma$ required for conformal Bjorken flow.  The energy conservation equation (\ref{eq:energydens}) can be expressed in terms of $\pi$ as
\be
\tau \partial_\tau  \! \log \epsilon  = -\frac{4}{3} + \frac{\pi}{\epsilon} \, ,
\label{eq:firstmom}
\ee 
where we have used $\pi = P_{\rm eq} - P_L$.
To relate $\pi$ and $\xi$ one can use this definition to obtain
\be
\overline\pi(\xi) \equiv \frac{\pi}{\epsilon} = \frac{1}{3} \left[ 1 - \frac{{\cal R}_L(\xi)}{{\cal R}(\xi)} \right]  .
\label{eq:pixirel}
\ee 
with
\be
{\cal R}_{L}(\xi) = \frac{3}{\xi} 
\left[ \frac{(\xi+1){\cal R}(\xi)-1}{\xi+1}\right] .
\ee

For conformal Bjorken flow, $\overline\pi$ is related to the shear inverse Reynolds number via  
\be
R_\pi^{-1} \equiv \frac{\sqrt{\pi^{\mu\nu} \pi_{\mu\nu}}}{P_{\rm eq}} = 3 \sqrt{\frac{3}{2}} |\overline\pi| \, .
\label{eq:reynoldsnumber}
\ee
As a consequence of Eq.~(\ref{eq:reynoldsnumber}), a series in $\overline\pi$ maps to a series in $R_\pi^{-1}$.

Changing variables to $\pi$ in \eqref{eq:2ndmomf} and using \eqref{eq:firstmom} one obtains \cite{Strickland:2017kux}
\be
\frac{\partial_\tau\pi}{\epsilon} + \frac{\pi}{\epsilon\tau} \left( \frac{4}{3} - \frac{\pi}{\epsilon}  \right) - \left[ \frac{2(1+\xi)}{\tau} - \frac{{\cal H}(\xi)}{\tau_{\rm eq}} \right]\overline\pi'(\xi) = 0\, ,
\label{eq:2ndmomf3}
\ee
where $\xi = \xi(\overline\pi)$ is the inverse function from $\overline\pi$ to $\xi$, $\overline\pi' \equiv d\overline\pi/d\xi$, and ${\cal H}(\xi) \equiv \xi (1+\xi)^{3/2}{\cal R}^{5/4}(\xi)$.
When expressed in this form one sees that the aHydro second-moment equation resums an infinite series in the inverse Reynolds number (\ref{eq:reynoldsnumber}).  This is because the last term on the right hand side of Eq.~(\ref{eq:2ndmomf3}) is a function that contains all orders in $\xi$ and, hence, $\overline\pi$.\footnote{A similar construction can be made in the case of Gubser flow, see Sec. IIIC of Ref.~\cite{Martinez:2017ibh}.}

Using small-anisotropy expansions one obtains~\cite{Strickland:2017kux}
\ba
\overline\pi^\prime  &=& \frac{8}{45} - \frac{26}{21} \overline\pi + \frac{1061}{392} \overline\pi^2 + {\cal O}(\overline\pi^3) \, , \nonumber \\
{\cal H} &=& \frac{45}{8} \overline\pi \left[ 1 + \frac{405}{56} \overline\pi + {\cal O}(\overline\pi^3) \right] .
\ea
Plugging these expansions into Eq.~(\ref{eq:2ndmomf3}) and keeping terms through second order in $\pi$ gives
\be
\partial_\tau \pi - \frac{4 \eta}{3 \tau_\pi \tau} + \frac{38}{21} \frac{\pi}{\tau} - \frac{36\tau_\pi}{245\eta} \frac{\pi^2}{\tau}
= - \frac{\pi}{\tau_\pi} - \frac{15}{56} \frac{\pi^2}{\tau_\pi \epsilon} +{\cal O}(\pi^3) \, .
\label{eq:2ndmomf4}
\ee
When truncated at linear order $\pi$, this evolution equation agrees exactly with previously obtained second-order viscous hydrodynamics evolution equations in relaxation time approximation \cite{Denicol:2010xn,Denicol:2012cn,Denicol:2014loa,Jaiswal:2013vta,Jaiswal:2013npa}.  This demonstrates that, in the limit of small momentum-space anistropy, aHydro automatically reproduces the correct second-order viscous hydrodynamics equations.  Note that it is possible to obtain higher-order terms such as those contributing at the order of the inverse Reynolds number squared as well.  When applied to phenomenology, one does not expand Eqs.~\eqref{eq:2ndmomf} or \eqref{eq:2ndmomf3} in $\bar\pi$ when solving the aHydro dynamical equations and an infinite number of orders in the inverse Reynolds number are automatically included.  This is why aHydro represents a {\em resummed dissipative hydrodynamic theory}.  In practice, aHydro automatically regulates the magnitude of $\bar\pi$ such that unphysical behaviour of the kinetic pressures, e.g., $P_L <0$, simply cannot occur.

%%%%%%%%%%%%%%%%%%%%%%%%%%%%%%%%%%%%%%%%%%
\section{Quasiparticle anisotropic hydrodynamics}
\label{sec:aHydroQP}
%%%%%%%%%%%%%%%%%%%%%%%%%%%%%%%%%%%%%%%%%%

In order to faithfully model heavy-ion collisions one must obtain the evolution equations for arbitrary 3+1D configurations and include the non-conformality of QCD consistent with a realistic lattice-based equation of state. In order to do this in {\em quasiparticle anisotropic hydrodynamics} we assume a system of massive relativistic quasiparticles with temperature-dependent masses $m(T)$. The system is assumed to obey a relativistic Boltzmann equation with $m(T)$ determined from lattice QCD (LQCD) computations of QCD thermodynamics. When the masses are temperature dependent, the Boltzmann equation contains an additional force term on the left-hand side related to gradients in the temperature, and hence gradients in $m$,
\be
p^\mu \partial_\mu f + \frac{1}{2}\partial_i m^2 \partial_{(p)}^{i} f =  \underbrace{- \frac{p \cdot u}{\tau_{\rm eq}(T)} [ f - f_\text{eq}(T) ] }_{C[f]}\, .
\label{eq:be}
\ee
The right-hand side of the Boltzmann equation is the collisional kernel $C[f]$ which we take to be given by the relaxation time approximation (RTA), where $u^\mu$ is the four-velocity associated with the local rest frame (LRF) of the matter and Latin indices such as $i$ indicate spatial indices. The collisional kernel is a functional of the one-particle distribution function $f(x,p)$ which depends on space-time coordinates $x$ and momentum $p$.
For a gas of massive quasiparticles, the relaxation time is given by \cite{Alqahtani:2017mhy}
\be{}\tau_{\rm eq}(T)= \bar{\eta} \, \frac{\epsilon+P}{I_{3,2}(\hat{m}_{\rm eq})}
\ee
where $\bar\eta = \eta/s$ is the specfic shear viscosity, $\epsilon$ is the energy density, $P$ is the pressure which fixed by the equation of state, and the special functions appearing are given by
\ba 
I_{3,2}(\hat{m}_{\rm eq}) &=& \frac{N_{\rm dof} T^5\, \hat{m}_{\rm eq}^5}{30 \pi^2} \bigg[ \frac{1}{16} \Big(K_5(\hat{m}_{\rm eq})-7K_3(\hat{m}_{\rm eq})+22 K_1(\hat{m}_{\rm eq}) \Big)-K_{i,1}(\hat{m}_{\rm eq}) \bigg] \, ,  \hspace{5mm} \\ 
K_{i,1}(\hat{m}_{\rm eq})&=&\frac{\pi}{2}\Big[1-\hat{m}_{\rm eq} K_0(\hat{m}_{\rm eq}) s_{-1}(\hat{m}_{\rm eq})-\hat{m}_{\rm eq}K_1(\hat{m}_{\rm eq}) s_0(\hat{m}_{\rm eq})\Big] \, ,
\ea
with $\hat{m}_{\rm eq}=m/T$, $N_{\rm dof}$ is being the number of degrees of freedom (degeneracy) , $K_n$ are the modified Bessel functions of the second kind, and $s_n$ are the modified Struve functions.
The effective temperature $T(\tau)$ is computed by requiring the non-equilibrium kinetic energy densities calculated from $f$ to be equal to the equilibrium kinetic energy density calculated from the equilibrium distribution, $f_\text{eq}(T,m)$. We note that the second term on the left-hand side of Boltzmann equation, $\frac{1}{2}\partial_i m^2 \partial_{(p)}^{i} f$, matches exactly the result obtained by deriving the Boltzmann equation using quantum field theoretical methods~\cite{2006}.

In this review, we assume the distribution function is given by the leading-order aHydro form, parameterized by a diagonal anisotropy tensor as follows
\be
f(x,p) = f_{\rm eq}\!\left(\frac{1}{\lambda} \sqrt{p_\mu \Xi^{\mu\nu} p_\nu} \right) 
\underset{{\rm LRF}}{\longrightarrow}   f_{\rm eq}\!\left(\frac{1}{\lambda}\sqrt{\sum_i \frac{p_i^2}{\alpha_i^2} + m^2}\right)  \, ,
\label{eq:fdef}
\ee
where $i\in \{x,y,z\}$, $\Xi^{\mu\nu}$ is the anisotropy tensor and the second equality holds in the LRF.  As indicated, in the LRF the argument of the distribution function can be expressed in terms of three independent momentum-anisotropy parameters $\alpha_i$.  Here we will assume that $f_{\rm eq}$ is given by a Boltzmann distribution which depends on $p \cdot u$ and the isotropic temperature T. Therefore, one can calculate the energy density and pressures by integrating the distribution function \eqref{eq:fdef} times $p^\mu p^\nu$ using the Lorentz-invariant integration measure $\int dP = \int \frac{d^3{\bf p}}{(2\pi)^3} \frac{1}{E}$. Performing the same operation allows one to extract all moments, and then one can create the requisite dynamical equations. 

The first aHydroQP equation of motion is obtained from the first moment of the left-hand side of the quasiparticle Boltzmann equation \eqref{eq:be}, which reduces to $\partial_\mu T^{\mu\nu}$. In the relaxation time approximation, however, the first moment of the collisional kernel on the right hand side results in a constraint that must be satisfied in order to conserve energy and momentum, i.e.$\int dP \, p^\mu C[f]=0$. This constraint can be enforced by expressing the effective temperature in terms of the microscopic parameters $\lambda$ and $\vec\alpha$.  As a consequence, computing the first moment of the Boltzmann equation gives the energy-momentum conservation law for the system
\be
\partial_\mu T^{\mu\nu}=0 \, ,
\label{eq:1m}
\ee
where %
\be
T^{\mu \nu}=\int \frac{d^3{\bf p}}{(2\pi)^3} \frac{1}{E}p^\mu p^\nu f \, .
\ee
%.
For the second equation of motion, we will perform a similar procedure using the second moment of the quasiparticle Boltzmann equation  
\be
\partial_\alpha  I^{\alpha\nu\lambda}- J^{(\nu} \partial^{\lambda)} m^2 =-\int \frac{d^3{\bf p}}{(2\pi)^3} \frac{1}{E} p^\nu p^\lambda{\cal C}[f]\, \label{eq:I-conservation} ,
\ee
with $I^{\mu\nu\lambda} \equiv \int \frac{d^3{\bf p}}{(2\pi)^3} \frac{1}{E}p^\mu p^\nu p^\lambda f $ and the the particle four-current $J^{\mu}=\int \frac{d^3{\bf p}}{(2\pi)^3} \frac{1}{E} p^\mu f$ . 

%----------------------------------------------------------------------------------------------------
\begin{figure}[t]
\centerline{\includegraphics[angle=0,width=0.35\textwidth]{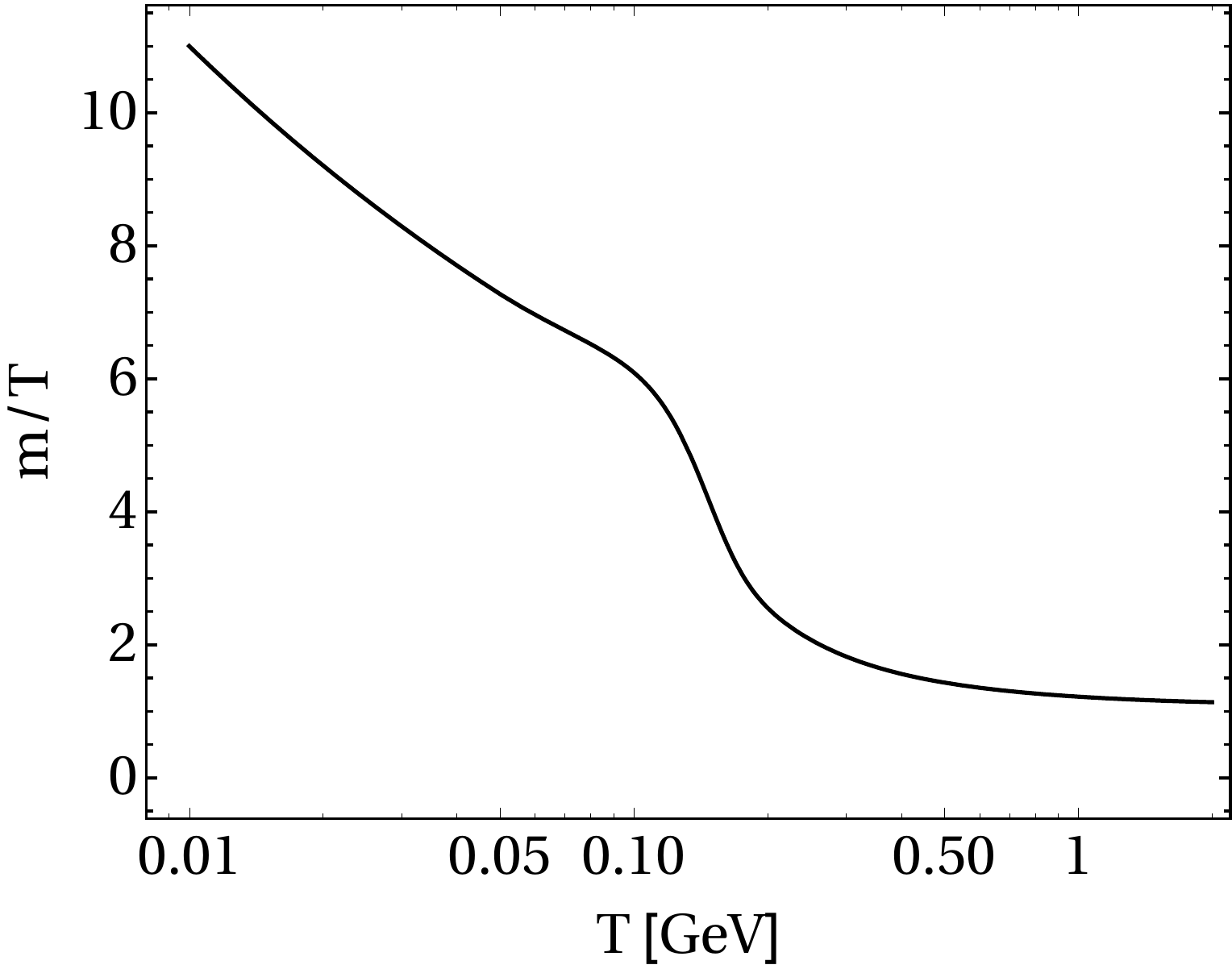}
\;\;
\includegraphics[angle=0,width=0.35\textwidth]{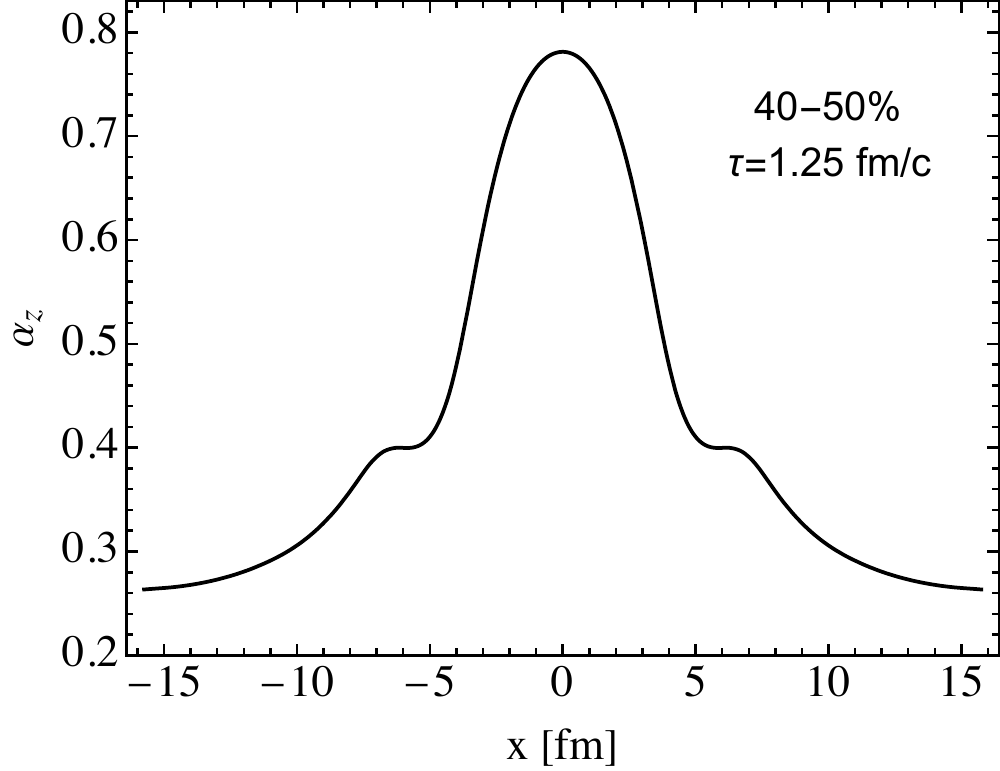}}
\caption{(a) The temperature dependence of the quasiparticle mass scaled by the temperature \cite{Alqahtani:2015qja}. (b) The
spatial profile of the anisotropy parameter $\alpha_z$ as a function of x \cite{Alqahtani:2020paa}.}
\label{fig:quasi_mass}
\end{figure}
%----------------------------------------------------------------------------------------------------

%%%%%%%%%%%%%%%%%%%%%%%%%%%%%%%%%%%%%%%%%%
\section{The equation of state for aHydroQP}
\label{sec:EoS}
%%%%%%%%%%%%%%%%%%%%%%%%%%%%%%%%%%%%%%%%%%

 For a system of massive particles obeying Boltzmann statistics, the equilibrium energy density, pressure, and entropy density are given by 
\ba
\epsilon_{\rm eq}(T,m) &=& 4 \pi \tilde{N} T^4 \, \hat{m}_{\rm eq}^2
 \Big[ 3 K_{2}\left( \hat{m}_{\rm eq} \right) + \hat{m}_{\rm eq} K_{1} \left( \hat{m}_{\rm eq} \right) \Big] \, , 
%\label{eq:EeqConstantM}
\\
 P_{\rm eq}(T,m) &=& 4 \pi \tilde{N} T^4 \, \hat{m}_{\rm eq}^2 K_2\left( \hat{m}_{\rm eq}\right)\, ,\\
 s_{\rm eq}(T,m) &=&4 \pi \tilde{N} T^3 \, \hat{m}_{\rm eq}^2 \Big[4K_2\left( \hat{m}_{\rm eq}\right)+\hat{m}_{\rm eq}K_1\left( \hat{m}_{\rm eq}\right)\Big] \, .
\label{eq:SeqConstantM}
\ea
In the quasiparticle approach, one assumes the mass to be temperature dependent, i.e. $m(T)$. This results in a change in the bulk variables Eqs.~\eqref{eq:SeqConstantM}. However, one can not simply insert $m(T)$ into the bulk variables since this will not be thermodynamically consistent. 
The entropy density may be obtained in two ways: $s_{\rm eq} = (\epsilon_{\rm eq} + P_{\rm eq})/T$ and $ s_{\rm eq} =\partial P_{\rm eq}/\partial T$. Then , by basically inserting a temperature-dependent mass m(T), the two identities will not give the same result. Therefore, the energy-momentum tensor definition needs a background field to correct this, i.e.
\be
T^{\mu\nu} = T^{\mu\nu}_{\rm kinetic} + g^{\mu\nu} B(T)  \, .
\label{eq:Tmunu}
\ee
where $B(T)$ is the additional background contribution. 
Thus, in an equilibrium Boltzmann gas with quasiparticles, the bulk thermodynamic variables for the gas become
\ba
\epsilon_{\rm eq}(T,m) &=& \epsilon_{\rm kinetic} +B_{\rm eq} \, , 
\label{eq:Eeq} \\
 P_{\rm eq}(T,m) &=& P_{\rm kinetic} -B_{\rm eq}\, ,\\
 \label{eq:Peq}
 s_{\rm eq}(T,m) &=&s_{\rm kinetic} \, .
\label{eq:Seq}
\ea
As a result of introducing the background field, the energy density and the pressure are modified by $+B_{\rm eq}$ and $-B_{\rm eq}$ terms, respectively.

 To determine the temperature-dependence of $B_{\rm eq}$ one requires thermodynamic consistency
 \be
 s_{\rm eq} = \epsilon_{\rm eq} + P_{\rm eq} = T \frac{\partial P_{\rm eq}}{\partial T}
 \ee
 However, we need to know in advance $m(T)$ to determine $B(T)$ which can be determined using the following thermodynamic identity
 \be
\epsilon_{\rm eq} + P_{\rm eq} = T s_{\rm eq} =  4 \pi \tilde{N} T^4 \, \hat{m}_{\rm eq}^3 K_3\left( \hat{m}_{\rm eq}\right)\,
 \ee
As we can see, one can solve numerically for $m(T)$ once the equilibrium energy density and pressure are determined  using the lattice QCD parameterization. The resulting effective mass scaled by $T$ extracted from continuum extrapolated Wuppertal-Budapest lattice data \cite{Borsanyi:2010cj} is shown in Fig.~\ref{fig:quasi_mass} (left panel) \cite{Alqahtani:2015qja}.   At high temperatures ($T \sim$ 0.6 GeV) the scaled mass is $ \sim T$ in agreement with the expected high-temperature behavior of QCD~\cite{Ryblewski:2017ybw}.

%%%%%%%%%%%%%%%%%%%%%%%%%%%%%%%%%%%%%%%%%%
\section{Evolution and freezeout in aHydroQP}
\label{sec:freezeout}
%%%%%%%%%%%%%%%%%%%%%%%%%%%%%%%%%%%%%%%%%%
The evolution equations for $u^\mu$, $\lambda$, and $\alpha_i$ are obtained from moments of the quasiparticle Boltzmann equation.  These can be expressed compactly by introducing a timelike vector $u^\mu$ which is normalized as $u^\mu u_\mu = 1$ and three spacelike vectors $X_i^\mu$ which are individually normalized as $X^\mu_i X_{\mu,i} = -1$.  These vectors are mutually orthogonal and obey $u_\mu X^\mu_i$ and $X_{\mu,i} X^\mu_j = 0$ for $i \neq j$ \cite{Ryblewski:2010ch,Martinez:2012tu}.  The four equations resulting from the first moment are
\ba
D_u\epsilon +\epsilon \theta_u + \sum_j P_j u_\mu D_j X^\mu_j &=&0\, , \label{eq:1stmomOne} \\
D_i P_i+P_i\theta_i -\epsilon X_{\mu,i} D_uu^\mu + P_i X_{\mu,i} D_i X^\mu_i - \sum_j P_j X_{\mu,i} D_j X^\mu_j  &=& 0\,, 
\label{eq:1stmomTwo} 
\ea
where $i,j \in \{x,y,z\}$, $D_u \equiv u^\mu \partial_\mu$, and $D_i \equiv X^\mu_i \partial_\mu$.  The expansion scalars are $\theta_u = \partial_\mu u^\mu$ and $\theta_i = \partial_\mu X^\mu_i$.  Explicit expressions for the basis vectors, derivative operators and expansion scalars can be found in Refs.~\cite{Nopoush:2014pfa,Alqahtani:2015qja,Alqahtani:2016rth,Alqahtani:2017tnq}.  The quantities $\epsilon$ and $P_i$ are the kinetic energy density and pressures obtained using the anisotropic hydrodynamics ansatz for the one-particle distributions function corrected by the background contribution $B(T)$ necessary to enforce thermodynamic consistency
\ba
\epsilon &=& \epsilon_{\rm kinetic}(\lambda,\vec\alpha,m) + B(\lambda,\vec\alpha) \, ,  \\
P_i &=& P_{i, \rm kinetic}(\lambda,\vec\alpha,m) - B(\lambda,\vec\alpha) \ \, ,
\ea

The three equations resulting from the second moment of the Boltzmann equation are
\be
D_u I_i + I_i (\theta_u + 2 u_\mu D_i X_i^\mu)
= \frac{1}{\tau_{\rm eq}} \Big[ I_{\rm eq}(T,m) - I_i \Big] ,
\label{eq:2ndmoment} 
\ee
with \cite{Nopoush:2014pfa}
\ba
I_i &=& \alpha \, \alpha_i^2 \, I_{\rm eq}(\lambda,m) \, , \nonumber \\ 
I_{\rm eq}(\lambda,m) &=&  4 \pi {\tilde N} \lambda^5 \hat{m}^3 K_3(\hat{m}) \, ,
\ea
where $\hat{m} = m/\lambda$, $\alpha = \alpha_x \alpha_y \alpha_z$ and $\tilde N = N_{\rm dof}/(2\pi)^3$, with $N_{\rm dof}$ being the number of degrees of freedom present in the theory under consideration.

Equations~\eqref{eq:1stmomOne}, \eqref{eq:1stmomTwo}, and \eqref{eq:2ndmoment} provide seven partial differential equations for $\vec{u}$, $\vec{\alpha}$, and $\lambda$ which we solve numerically.  To determine the local effective temperature we make use of Landau matching; requiring the equilibrium and non-equilibrium energy densities in the LRF to be equal and solving for $T$. Herein, we assume the system to initially be isotropic in momentum space $\alpha_i(\tau_0)=1$, with zero transverse flow. However, the system evolves quite fast away from isotropy $\tau_{\rm aniso} \lesssim 1$ fm. As an example, in Fig.~\ref{fig:quasi_mass}-right panel we show the spatial profile of the longitudinal anisotropy parameter at 40-50\% centrality class. As can be seen from this figure, $\alpha_z$ differs from unity especially in the dilute regions $|x|>5$ fm. We note here that no regulation is required in aHydroQP to evolve in these dilute regions. This system of partial differential equations keep evolving until the effective temperature in the entire simulation volume falls below a given freeze-out temperature of $T_{\rm FO}$.  From the results, we extract a three-dimensional freeze-out hypersurface with a fixed energy density (temperature). We assume in this step that the fluid anisotropy tensor and scale parameter are the same for all hadronic species.  We also assume that all hadrons created are in chemical equilibrium. With the use of an extended Cooper-Frye prescription \cite{Alqahtani:2017mhy}, we are able to translate the underlying hydrodynamic evolution values for the flow velocity, the anisotropy parameters, and the scale into explicit 'primordial' hadronic distribution functions on this hypersurface. 

The values of the aHydroQP parameters on the freezeout hypersurface are passed to a modified version of THERMINATOR 2 \cite{Chojnacki:2011hb}, which generates hadronic configurations using Monte-Carlo sampling. After sampling the primordial hadrons, further hadronic decays are taken into account using the built-in routines in THERMINATOR 2. The source code for aHydroQP and our custom version of THERMINATOR 2 are both freely accessible \cite{kent-code-library}. The aHydroQP formalism was used at different collision energies, and it was found that the observed differential spectra of identified hadrons, charged particle multiplicity, elliptic flow, and Hanbury-Brown-Twiss radii could be reproduced. 

Finally, in Table \ref{table:1} we list the fitting parameters that we extracted and used in the comparisons.
\begin{table}[h!]
\centering
\begin{tabular}{ |c|c|c|c| } 
\hline
 collision energy & $T_{0}$ [MeV] & $\eta/s$ %& {$T_{\rm FO}$ [MeV]}
 \\
 \hline
 200 GeV  & 455  & 0.179 %& 130}
 \\
\hline
 2.76 TeV  & 600  & 0.159 %& 130}
 \\
 \hline
 5.02 TeV  & 630  & 0.159 %& 130}
 \\
 \hline
\end{tabular}
\caption{The key parameters used in the presented results. }
\label{table:1}
\end{table}
%

%%%%%%%%%%%%%%%%%%%%%%%%%%%%%%%%%%%%%%%%%%
\section{Results and Discussion}
\label{sec:Results}
%%%%%%%%%%%%%%%%%%%%%%%%%%%%%%%%%%%%%%%%%%
In this section we present phenomenological comparisons of 3+1D aHydroQP model to experimental data. For the sake of brevity, we present comparisons of a small set of observables performed at various collision energies $\sqrt{s_{NN}}$ = 200 GeV, 2.76 , 5.02 TeV for Au-Au and Pb-Pb collisions from the PHENIX, PHOBOS, STAR, and ALICE collaborations.

%----------------------------------------------------------------------------------------------------
\begin{figure}[t]
\centerline{
\includegraphics[angle=0,width=0.35\textwidth]{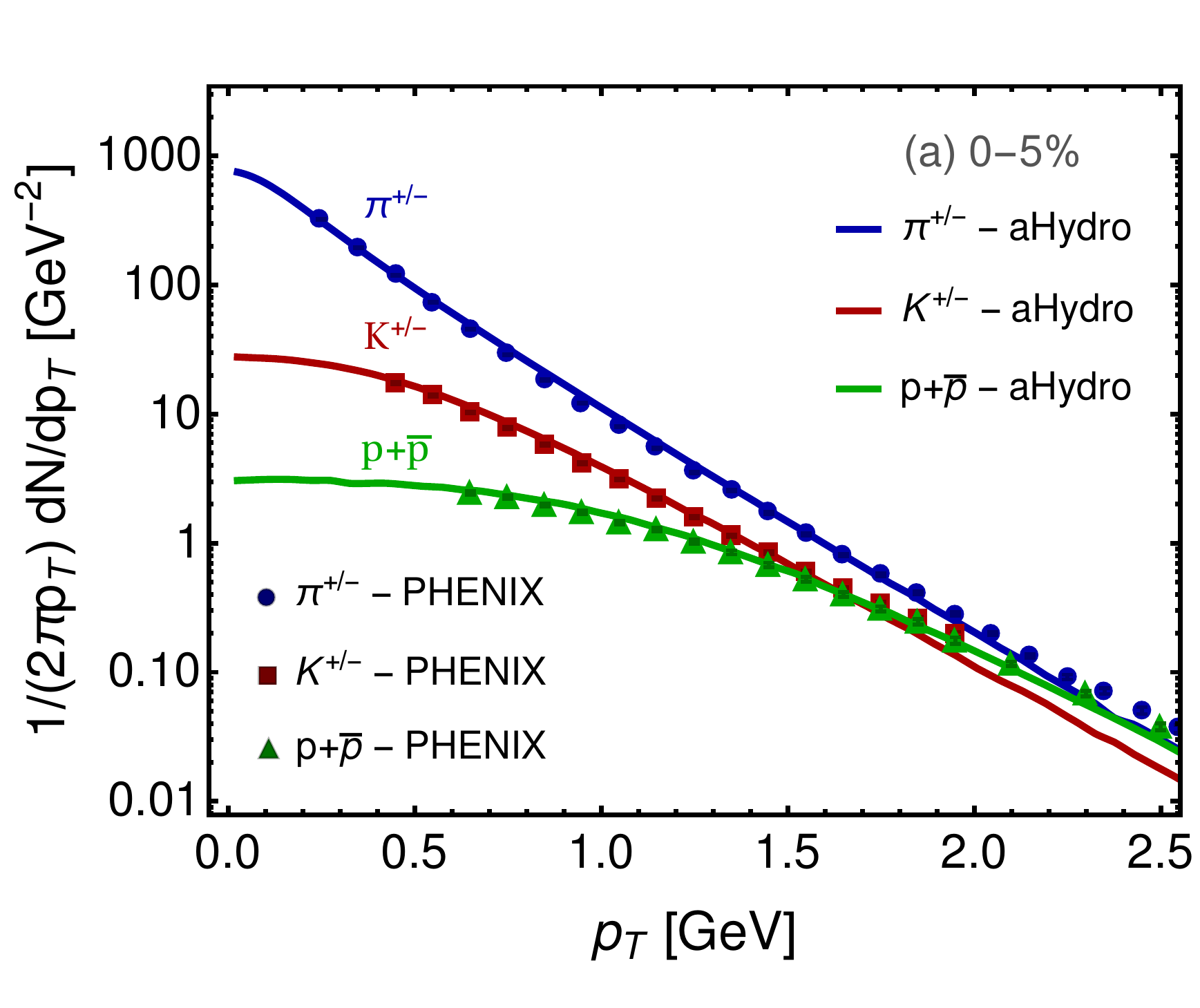} \;\;
\includegraphics[angle=0,width=0.35\textwidth]{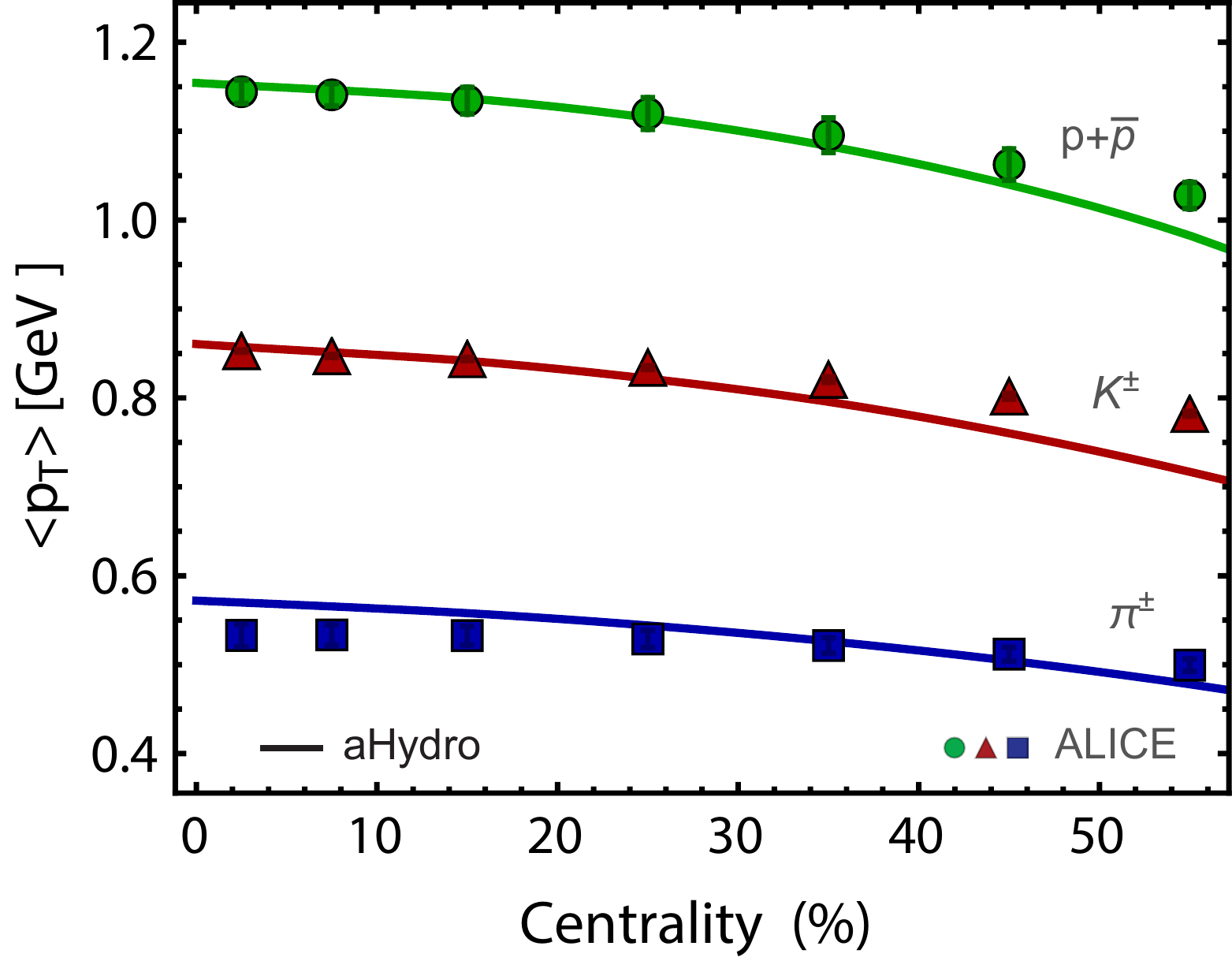}}
\caption{Left: Pion, kaon, and proton spectra compared to experimental data by the PHENIX collaboration at 200 GeV for Au-Au collisions \cite{PHENIX:2003iij,Almaalol:2018gjh}. 
Right: Pion, kaon, and proton average transverse momentum as a function of  centrality compared to data by the ALICE collaboration at 2.76 TeV for Pb-Pb collisions \cite{ALICE:2013mez,Alqahtani:2017tnq}}
\label{fig:ptc}
\end{figure}
%----------------------------------------------------------------------------------------------------
We first present comparisons of bulk observables between our model and experimental results. In Fig. ~\ref{fig:ptc}-left panel, we show the spectra of pions, kaons, and protons as a function of the transverse momentum $p_T$. The agreement shown between our model and the experimental results is good up to quite large $p_T \sim 2$ GeV. In this figure, we show only one centrality class 0-5\%, however one can see Ref. \cite{Almaalol:2018gjh} for more comparisons up to 30-40\% centrality class. It suffices here to say that the agreement is quite good up to $p_T \sim 1.5$ GeV for high centrality classes. Next, we present the centrality dependence of the average transverse momentum of pions, kaons, and protons at 2.76 TeV for Pb-Pb collisions. Again, the agreement is very good up to high centrality classes $ \sim 50$\%. The spectra at this energy is not presented here, however, it can be found in Ref.\cite{Alqahtani:2017tnq}, where the agreement between aHydroQP model and the data for different centrality classes is good.

%----------------------------------------------------------------------------------------------------
\begin{figure}[t]
\centerline{
\includegraphics[angle=0,width=0.33\textwidth]{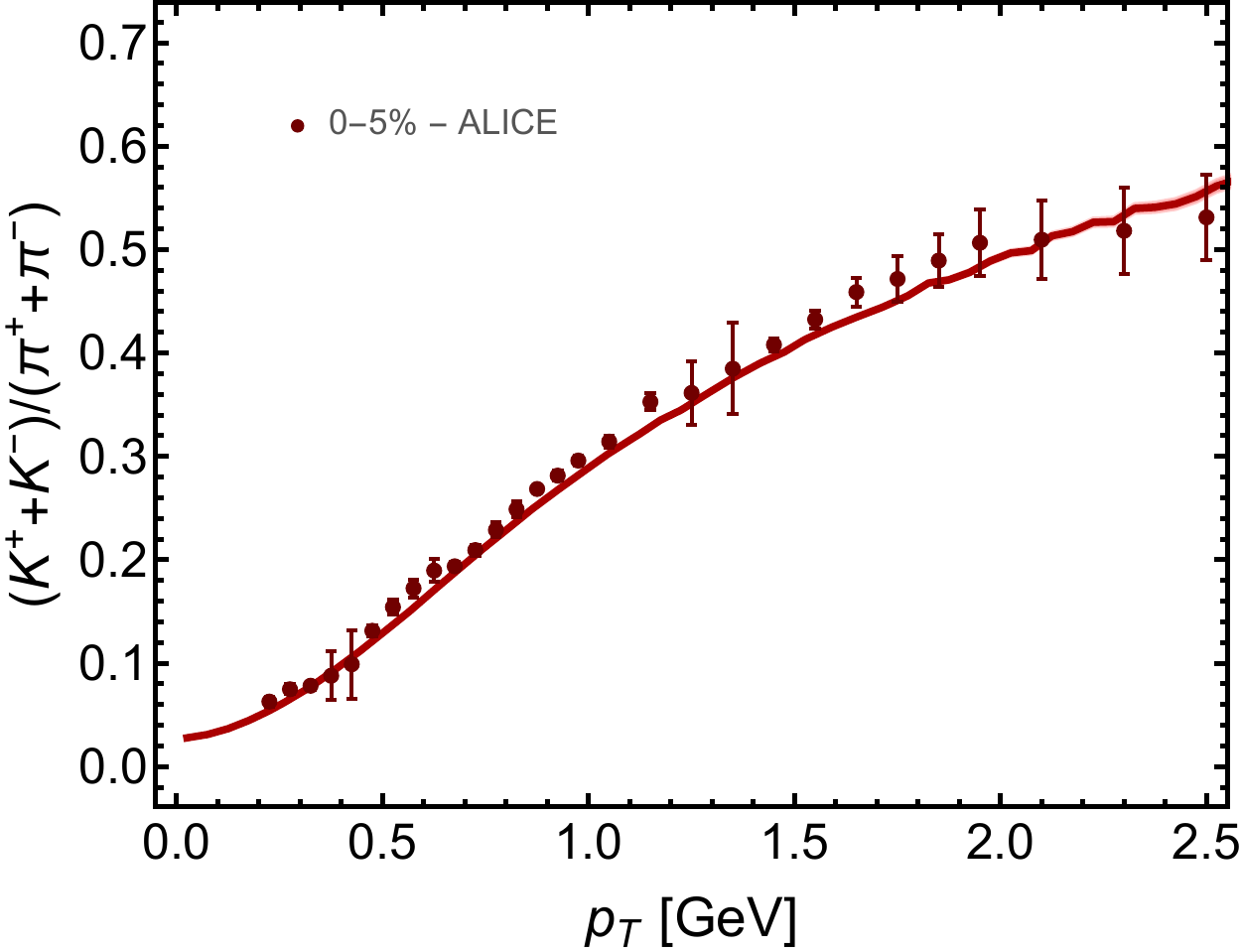}
\;\;
\includegraphics[angle=0,width=0.35\textwidth]{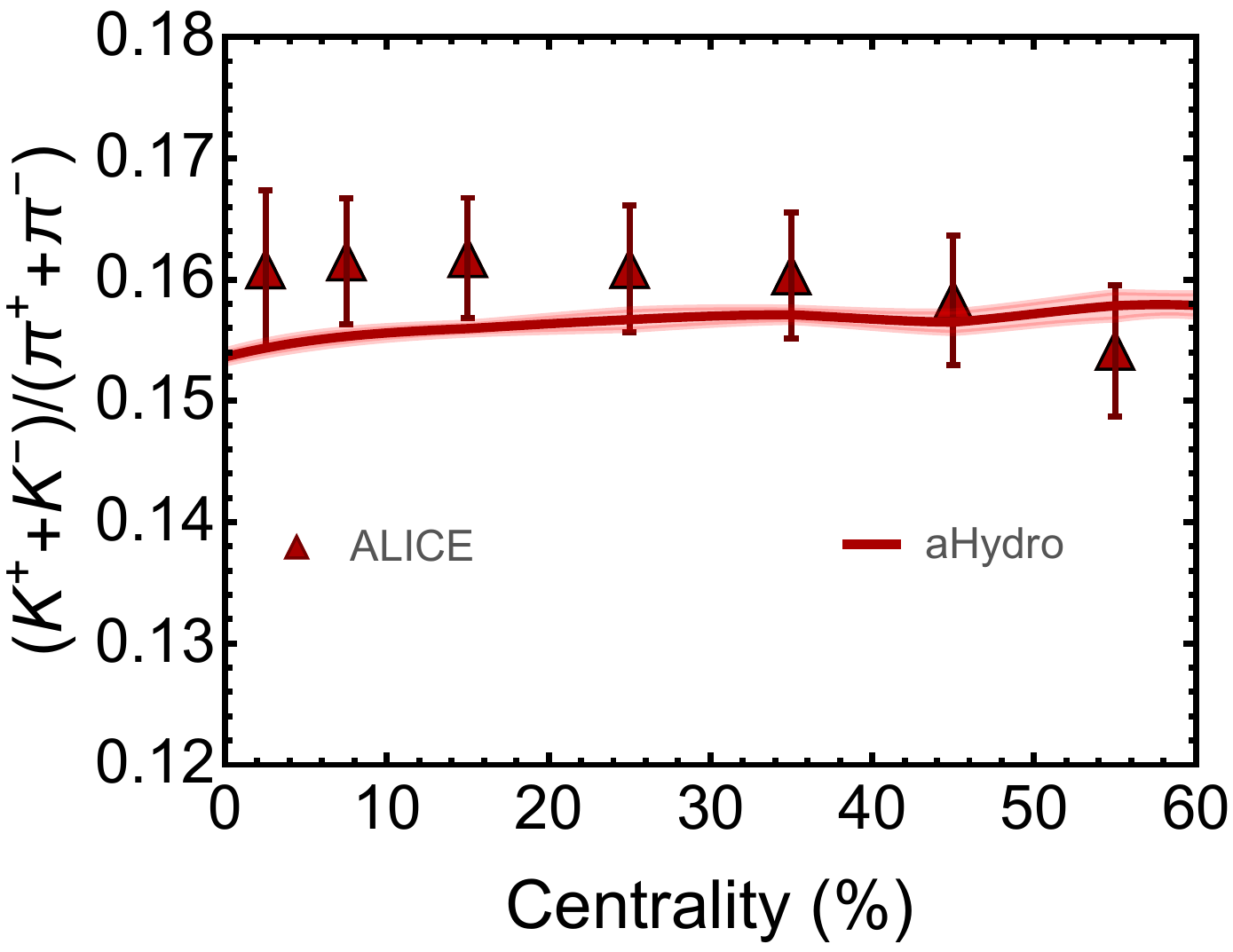}}
\caption{left: The kaon-to-pion ratio as a function of the transverse momentum. Right: The centrality dependence of the kaon-to-pion ratio.  In both panels, the predictions of aHydroQP model are compared to experimental data from the ALICE collaboration in Pb-Pb collisions at $\sqrt{s_{NN}}$ = 5.02 TeV.  \cite{ALICE:2019hno,Alqahtani:2020paa}. }
\label{fig:kpc}
\end{figure}
%----------------------------------------------------------------------------------------------------
Next, in the left panel of Fig.~\ref{fig:kpc}, the kaon-to-pion ratio $(K^+ + K^-)/ (\pi^+ +\pi^-)$ is presented as a function of $p_T$ in the 0-5\% centrality class. As can be seen from this figure, our model was able to reproduce the ratios well up to fairly large $p_T \sim 2.5$ GeV. We also show, in Fig.~\ref{fig:kpc}-right panel, the kaon-to-pion ratio as a function of centrality where our model again describes the data quite well over a wide range of centrality classes. In ref.\cite{Alqahtani:2020paa}, we showed the kaon-to-pion ratio and also the proton-to-pion ratio for multiple different centrality classes with a reasonable agreement to the data at 5.023 TeV for Pb-Pb collisions from ALICE collaboration.

%----------------------------------------------------------------------------------------------------
\begin{figure}[t]
\centerline{
\includegraphics[angle=0,width=0.35\textwidth]{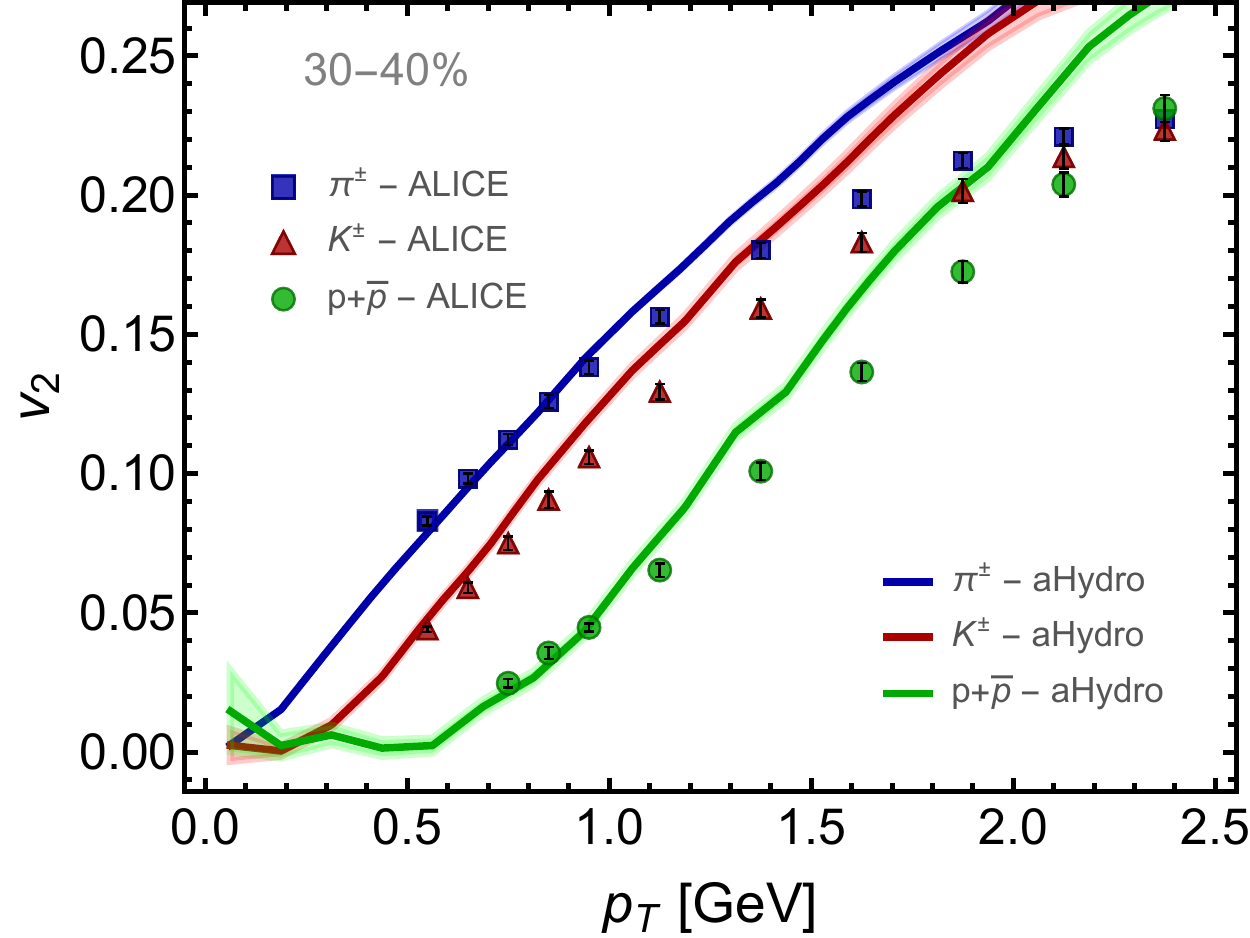}
\;\;
\includegraphics[angle=0,width=0.35\textwidth]{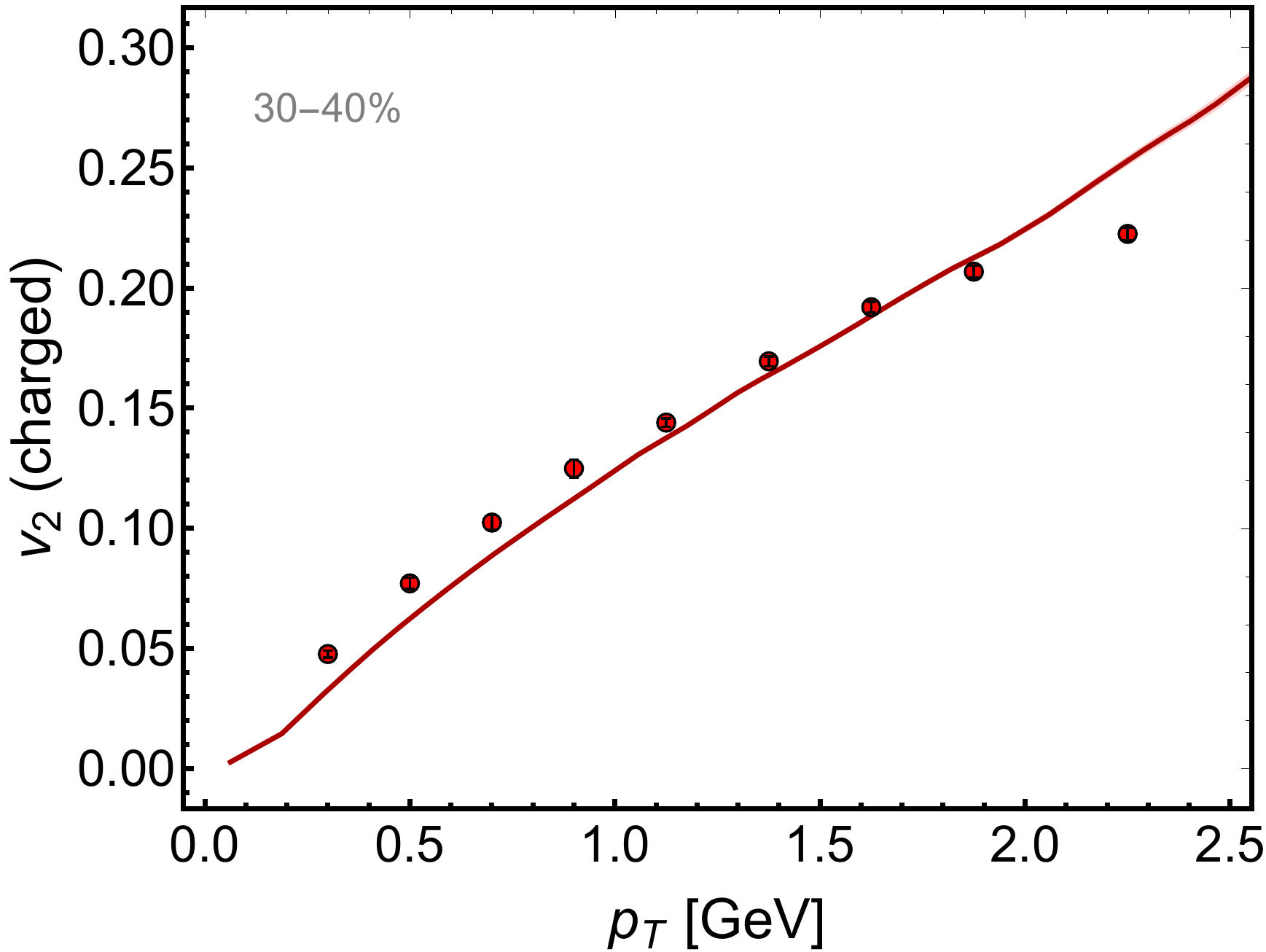}}
\caption{Left: The elliptic flow coefficient ($v_2$) as a function of the transverse momentum \cite{ALICE:2018yph}. Right: The $p_T$ dependence of $v_2$ of all charged particles \cite{ALICE:2016ccg}. In both panels, aHydroQP predictions are compared to experimental data from the ALICE collaboration in Pb-Pb collisions at $\sqrt{s_{NN}}$ = 5.02 TeV in the 30-40\% centrality class \cite{Alqahtani:2020paa}.}
\label{fig:v2pt}
\end{figure}
%----------------------------------------------------------------------------------------------------
Next, we present comparisons of the anisotropic flow at 5.023 TeV for Pb-Pb collisions from the ALICE collaboration. In Fig.~\ref{fig:v2pt}-left panel, the identified elliptic flow coefficient $v_2$ is shown in the 30-40\% centrality class. A similar agreement between this model and experimental data is seen across different energies, see Refs.\cite{Almaalol:2018gjh,Alqahtani:2017tnq,Alqahtani:2020paa}. Moreover, the elliptic flow for charged hadrons as a function of $p_T$ is shown in Fig.~\ref{fig:v2pt}-right panel at 30-40\% centrality class. One can see that our model prediction agrees quite well with the data up to $p_T$ GeV.

\begin{figure}[t]
\centerline{
\includegraphics[angle=0,width=0.35\textwidth]{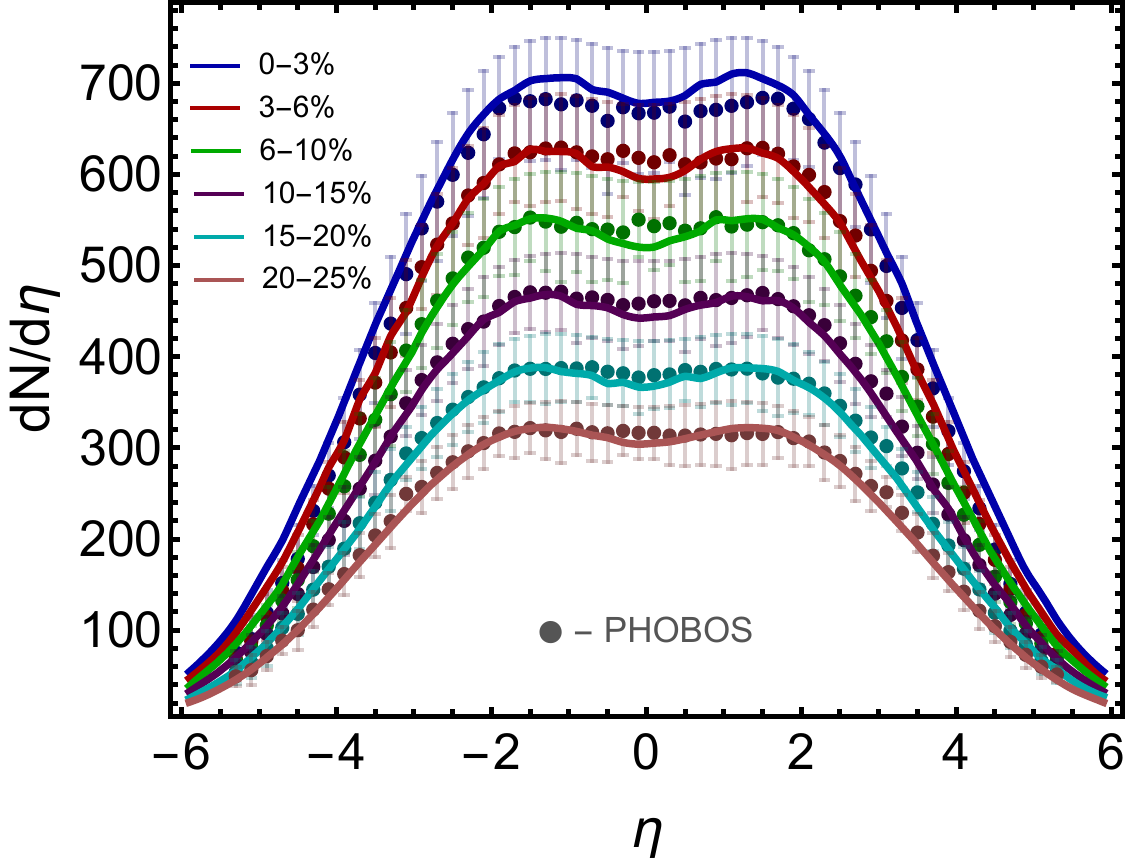}
\;\;
\includegraphics[angle=0,width=0.35\textwidth]{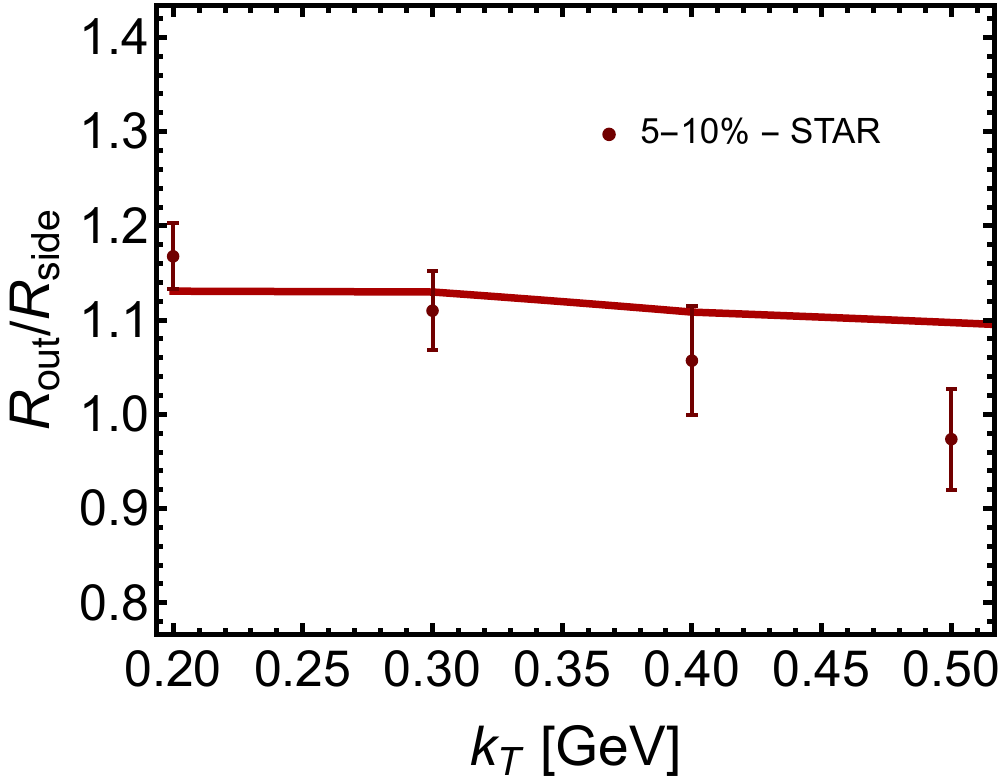}}
\caption{Left: Charged particle multiplicity as a function of pseudorapidity where data are from the PHOBOS collaboration \cite{PHOBOS:2010eyu,Almaalol:2018gjh}. 
Right: The $R_{\rm out}/R_{\rm side} $ ratio as a function of the pair mean transverse momentum ($k_{T}$) for $\pi^+\pi^+$ in the  5-10\% centrality class where data are from the STAR collaboration \cite{STAR:2004qya,Alqahtani:2020daq}. In both panels, results are from Au-Au collisions at $\sqrt{s_{NN}}$=200 GeV.}
\label{fig:qm_rhicmult}
\end{figure}

Furthermore, we present, in Fig. ~\ref{fig:qm_rhicmult}-left panel, the charged particle multiplicity as a function of the pseduorapidity where data are from the PHOBOS collaboration for Au+Au collisions at $\sqrt{s_{NN}}$=200 GeV. In this plot, the multiplicity is shown for different centrality classes: 0-3\%, 3-6\%, 6-10\%, 10-15\%, 15-20\%, and 20-25\%. We find that our model does a good job in reproducing the pseudorapidity dependence of the multiplicity in a wide range of centrality classes. A similar agreement to the data using aHydroQP model is observed for other systems at different energies \cite{Alqahtani:2020paa,Alqahtani:2017tnq}.

Finally, we present the aHydroQP predictions for HBT radii determined from pion correlations. As an example, in Fig. ~\ref{fig:qm_rhicmult}-right panel, we show the $R_{out}/R_{side}$ ratio as a function of the mean transverse momentum of the pair $\pi^+\pi^+$ in the 5-10 \% centrality class. As can be seen from this figure, our model was able to describe the experimental data from the STAR collaboration quite well especially for low $k_T$ up to $ \sim 0.4$ GeV. For more details, see \cite{Alqahtani:2020daq}. We not here that a similar agreement of the HBT radii and thier ratios to the data is seen at 2.76 TeV for Pb-Pb collisions \cite{Alqahtani:2017tnq}.

%----------------------------------------------------------------------------------------------------
%----------------------------------------------------------------------------------------------------

%%%%%%%%%%%%%%%%%%%%%%%%%%%%%%%%%%%%%%%%%%
\section{Conclusions}
\label{sec:conclusions}
%%%%%%%%%%%%%%%%%%%%%%%%%%%%%%%%%%%%%%%%%%

In this review we presented a summary of recent progress in anisotropic hydrodynamics and its application to heavy-ion phenomenology.  We began with a demonstration in the simple case of conformal Bjorken expansion that aHydro resums an infinite number of terms in the inverse Reynolds number.  This feature allows aHydro to better describe systems that are far from equilibrium than traditional approaches.  In AA collision large non-equilibrium corrections occur during the initial stages of the QGP ($\tau < 1$ fm/c) and at all times near the cold edges of the plasma where the relaxation time grows large.  In collisions of small systems such as pA and pp one expects that, if a QGP is generated, it will be much more short-lived than in central AA collisions due to larger transverse gradients and, as a consequence, it will experience larger deviations from equilibrium during its evolution and freeze out.

Turning to AA phenomenology, we presented comparisons between the 3+1D aHydroQP model and heavy-ion experimental data collected at RHIC and LHC.  We list the extracted initial central temperature and shear viscosity at 200 GeV, 2.76 TeV, and 5.02 TeV in Table \ref{table:1}.  At all three collision energies with these parameters, we were able to describe the identified hadron spectra well, including the $p_T$-dependence of the kaon to pion ratio.  In addition, the extracted integrated elliptic flow for charged particles and $p_T$-dependence of the pion, proton, and kaon elliptic flow were found to also be in good agreement with the data.  Finally, we also presented comparisons between aHydroQP model predictions and STAR data from the ratio of 'out' and 'side' HBT radii, again finding good agreement with the data given current experimental uncertainties.

Looking to the future, final work is underway to release a new computational pipeline for 3+1D aHydroQP which includes fluctuating initial conditions of various types such as Trento \cite{Moreland:2014oya,Ke:2016jrd} or IP-Glasma \cite{Bartels:2002cj,Kowalski:2003hm,Schenke:2012wb}, a custom anisotropic hadronic freeze-out sampler based on ISS \cite{iss-url}, and full URQMD \cite{Bass:1998ca,Bleicher:1999xi} or SMASH \cite{Weil:2016zrk} hadronic afterburners that include elastic as well as inelastic channels.  Once complete, this will allow us to compute higher-order flow coefficients using aHydroQP in AA, pA, and pp collisions. 

\authorcontributions{H. Alalawi, M. Alqahtani,  and M. Strickland contributed equally to writing this review.}

\funding{H. Alalawi was supported by a visiting Ph.D. scholarship from Umm Al-Qura University. M. Alqahtani was supported by the Deanship of Scientific Research at the Imam Abdulrahman Bin Faisal University under grant number 2021-089-CED. M. Strickland was supported by the U.S. Department of Energy, Office of Science, Office of Nuclear Physics under Award No. DE-SC0013470.}

\dataavailability{The code necessary to generate all results shown can be downloaded from \url{https://www.personal.kent.edu/\~mstrick6/code/index.html}.}

\conflictsofinterest{The authors declare no conflict of interest.} 

\end{paracol}

\reftitle{References}
\externalbibliography{yes}
\bibliography{aHydroRev}

\end{document}